\documentclass[sigconf]{acmart}

\usepackage{sc25repro}
\usepackage{algorithmic}
\usepackage{graphicx}
\usepackage{caption}
\usepackage{subcaption}
\usepackage{listings}
\usepackage{fancyvrb}
\usepackage{textcomp}
\usepackage{xcolor}
\usepackage{array}
\usepackage{makecell}
\usepackage{url}
\usepackage{enumitem}
\usepackage{multirow}

\AtBeginDocument{%
  }

\makeatletter
\def\PYG@reset{\let\PYG@it=\relax \let\PYG@bf=\relax%
    \let\PYG@ul=\relax \let\PYG@tc=\relax%
    \let\PYG@bc=\relax \let\PYG@ff=\relax}
\def\PYG@tok#1{\csname PYG@tok@#1\endcsname}
\def\PYG@toks#1+{\ifx\relax#1\empty\else%
    \PYG@tok{#1}\expandafter\PYG@toks\fi}
\def\PYG@do#1{\PYG@bc{\PYG@tc{\PYG@ul{%
    \PYG@it{\PYG@bf{\PYG@ff{#1}}}}}}}
\def\PYG#1#2{\PYG@reset\PYG@toks#1+\relax+\PYG@do{#2}}

\@namedef{PYG@tok@w}{\def\PYG@tc##1{\textcolor[rgb]{0.73,0.73,0.73}{##1}}}
\@namedef{PYG@tok@c}{\def\PYG@tc##1{\textcolor[rgb]{0.53,0.53,0.53}{##1}}}
\@namedef{PYG@tok@cp}{\let\PYG@bf=\textbf\def\PYG@tc##1{\textcolor[rgb]{0.80,0.00,0.00}{##1}}}
\@namedef{PYG@tok@cs}{\let\PYG@bf=\textbf\def\PYG@tc##1{\textcolor[rgb]{0.80,0.00,0.00}{##1}}\def\PYG@bc##1{{\setlength{\fboxsep}{0pt}\colorbox[rgb]{1.00,0.94,0.94}{\strut ##1}}}}
\@namedef{PYG@tok@s}{\def\PYG@tc##1{\textcolor[rgb]{0.87,0.13,0.00}{##1}}\def\PYG@bc##1{{\setlength{\fboxsep}{0pt}\colorbox[rgb]{1.00,0.94,0.94}{\strut ##1}}}}
\@namedef{PYG@tok@sr}{\def\PYG@tc##1{\textcolor[rgb]{0.00,0.53,0.00}{##1}}\def\PYG@bc##1{{\setlength{\fboxsep}{0pt}\colorbox[rgb]{1.00,0.94,1.00}{\strut ##1}}}}
\@namedef{PYG@tok@sx}{\def\PYG@tc##1{\textcolor[rgb]{0.13,0.73,0.13}{##1}}\def\PYG@bc##1{{\setlength{\fboxsep}{0pt}\colorbox[rgb]{0.94,1.00,0.94}{\strut ##1}}}}
\@namedef{PYG@tok@ss}{\def\PYG@tc##1{\textcolor[rgb]{0.67,0.40,0.00}{##1}}\def\PYG@bc##1{{\setlength{\fboxsep}{0pt}\colorbox[rgb]{1.00,0.94,0.94}{\strut ##1}}}}
\@namedef{PYG@tok@si}{\def\PYG@tc##1{\textcolor[rgb]{0.20,0.20,0.73}{##1}}\def\PYG@bc##1{{\setlength{\fboxsep}{0pt}\colorbox[rgb]{1.00,0.94,0.94}{\strut ##1}}}}
\@namedef{PYG@tok@se}{\def\PYG@tc##1{\textcolor[rgb]{0.00,0.27,0.87}{##1}}\def\PYG@bc##1{{\setlength{\fboxsep}{0pt}\colorbox[rgb]{1.00,0.94,0.94}{\strut ##1}}}}
\@namedef{PYG@tok@ow}{\def\PYG@tc##1{\textcolor[rgb]{0.00,0.53,0.00}{##1}}}
\@namedef{PYG@tok@k}{\let\PYG@bf=\textbf\def\PYG@tc##1{\textcolor[rgb]{0.00,0.53,0.00}{##1}}}
\@namedef{PYG@tok@kp}{\def\PYG@tc##1{\textcolor[rgb]{0.00,0.53,0.00}{##1}}}
\@namedef{PYG@tok@kt}{\let\PYG@bf=\textbf\def\PYG@tc##1{\textcolor[rgb]{0.53,0.53,0.53}{##1}}}
\@namedef{PYG@tok@nc}{\let\PYG@bf=\textbf\def\PYG@tc##1{\textcolor[rgb]{0.73,0.00,0.40}{##1}}}
\@namedef{PYG@tok@ne}{\let\PYG@bf=\textbf\def\PYG@tc##1{\textcolor[rgb]{0.73,0.00,0.40}{##1}}}
\@namedef{PYG@tok@nf}{\let\PYG@bf=\textbf\def\PYG@tc##1{\textcolor[rgb]{0.00,0.40,0.73}{##1}}}
\@namedef{PYG@tok@py}{\let\PYG@bf=\textbf\def\PYG@tc##1{\textcolor[rgb]{0.20,0.40,0.60}{##1}}}
\@namedef{PYG@tok@nn}{\let\PYG@bf=\textbf\def\PYG@tc##1{\textcolor[rgb]{0.73,0.00,0.40}{##1}}}
\@namedef{PYG@tok@nb}{\def\PYG@tc##1{\textcolor[rgb]{0.00,0.20,0.53}{##1}}}
\@namedef{PYG@tok@nv}{\def\PYG@tc##1{\textcolor[rgb]{0.20,0.40,0.60}{##1}}}
\@namedef{PYG@tok@vc}{\def\PYG@tc##1{\textcolor[rgb]{0.20,0.40,0.60}{##1}}}
\@namedef{PYG@tok@vi}{\def\PYG@tc##1{\textcolor[rgb]{0.20,0.20,0.73}{##1}}}
\@namedef{PYG@tok@vg}{\def\PYG@tc##1{\textcolor[rgb]{0.87,0.47,0.00}{##1}}}
\@namedef{PYG@tok@no}{\let\PYG@bf=\textbf\def\PYG@tc##1{\textcolor[rgb]{0.00,0.20,0.40}{##1}}}
\@namedef{PYG@tok@nt}{\let\PYG@bf=\textbf\def\PYG@tc##1{\textcolor[rgb]{0.73,0.00,0.40}{##1}}}
\@namedef{PYG@tok@na}{\def\PYG@tc##1{\textcolor[rgb]{0.20,0.40,0.60}{##1}}}
\@namedef{PYG@tok@nd}{\def\PYG@tc##1{\textcolor[rgb]{0.33,0.33,0.33}{##1}}}
\@namedef{PYG@tok@nl}{\let\PYG@it=\textit\def\PYG@tc##1{\textcolor[rgb]{0.20,0.40,0.60}{##1}}}
\@namedef{PYG@tok@m}{\let\PYG@bf=\textbf\def\PYG@tc##1{\textcolor[rgb]{0.00,0.00,0.87}{##1}}}
\@namedef{PYG@tok@gh}{\def\PYG@tc##1{\textcolor[rgb]{0.20,0.20,0.20}{##1}}}
\@namedef{PYG@tok@gu}{\def\PYG@tc##1{\textcolor[rgb]{0.40,0.40,0.40}{##1}}}
\@namedef{PYG@tok@gd}{\def\PYG@tc##1{\textcolor[rgb]{0.00,0.00,0.00}{##1}}\def\PYG@bc##1{{\setlength{\fboxsep}{0pt}\colorbox[rgb]{1.00,0.87,0.87}{\strut ##1}}}}
\@namedef{PYG@tok@gi}{\def\PYG@tc##1{\textcolor[rgb]{0.00,0.00,0.00}{##1}}\def\PYG@bc##1{{\setlength{\fboxsep}{0pt}\colorbox[rgb]{0.87,1.00,0.87}{\strut ##1}}}}
\@namedef{PYG@tok@gr}{\def\PYG@tc##1{\textcolor[rgb]{0.67,0.00,0.00}{##1}}}
\@namedef{PYG@tok@ge}{\let\PYG@it=\textit}
\@namedef{PYG@tok@gs}{\let\PYG@bf=\textbf}
\@namedef{PYG@tok@gp}{\def\PYG@tc##1{\textcolor[rgb]{0.33,0.33,0.33}{##1}}}
\@namedef{PYG@tok@go}{\def\PYG@tc##1{\textcolor[rgb]{0.53,0.53,0.53}{##1}}}
\@namedef{PYG@tok@gt}{\def\PYG@tc##1{\textcolor[rgb]{0.67,0.00,0.00}{##1}}}
\@namedef{PYG@tok@err}{\def\PYG@tc##1{\textcolor[rgb]{0.65,0.09,0.09}{##1}}\def\PYG@bc##1{{\setlength{\fboxsep}{0pt}\colorbox[rgb]{0.89,0.82,0.82}{\strut ##1}}}}
\@namedef{PYG@tok@kc}{\let\PYG@bf=\textbf\def\PYG@tc##1{\textcolor[rgb]{0.00,0.53,0.00}{##1}}}
\@namedef{PYG@tok@kd}{\let\PYG@bf=\textbf\def\PYG@tc##1{\textcolor[rgb]{0.00,0.53,0.00}{##1}}}
\@namedef{PYG@tok@kn}{\let\PYG@bf=\textbf\def\PYG@tc##1{\textcolor[rgb]{0.00,0.53,0.00}{##1}}}
\@namedef{PYG@tok@kr}{\let\PYG@bf=\textbf\def\PYG@tc##1{\textcolor[rgb]{0.00,0.53,0.00}{##1}}}
\@namedef{PYG@tok@bp}{\def\PYG@tc##1{\textcolor[rgb]{0.00,0.20,0.53}{##1}}}
\@namedef{PYG@tok@fm}{\let\PYG@bf=\textbf\def\PYG@tc##1{\textcolor[rgb]{0.00,0.40,0.73}{##1}}}
\@namedef{PYG@tok@vm}{\def\PYG@tc##1{\textcolor[rgb]{0.20,0.40,0.60}{##1}}}
\@namedef{PYG@tok@sa}{\def\PYG@tc##1{\textcolor[rgb]{0.87,0.13,0.00}{##1}}\def\PYG@bc##1{{\setlength{\fboxsep}{0pt}\colorbox[rgb]{1.00,0.94,0.94}{\strut ##1}}}}
\@namedef{PYG@tok@sb}{\def\PYG@tc##1{\textcolor[rgb]{0.87,0.13,0.00}{##1}}\def\PYG@bc##1{{\setlength{\fboxsep}{0pt}\colorbox[rgb]{1.00,0.94,0.94}{\strut ##1}}}}
\@namedef{PYG@tok@sc}{\def\PYG@tc##1{\textcolor[rgb]{0.87,0.13,0.00}{##1}}\def\PYG@bc##1{{\setlength{\fboxsep}{0pt}\colorbox[rgb]{1.00,0.94,0.94}{\strut ##1}}}}
\@namedef{PYG@tok@dl}{\def\PYG@tc##1{\textcolor[rgb]{0.87,0.13,0.00}{##1}}\def\PYG@bc##1{{\setlength{\fboxsep}{0pt}\colorbox[rgb]{1.00,0.94,0.94}{\strut ##1}}}}
\@namedef{PYG@tok@sd}{\def\PYG@tc##1{\textcolor[rgb]{0.87,0.13,0.00}{##1}}\def\PYG@bc##1{{\setlength{\fboxsep}{0pt}\colorbox[rgb]{1.00,0.94,0.94}{\strut ##1}}}}
\@namedef{PYG@tok@s2}{\def\PYG@tc##1{\textcolor[rgb]{0.87,0.13,0.00}{##1}}\def\PYG@bc##1{{\setlength{\fboxsep}{0pt}\colorbox[rgb]{1.00,0.94,0.94}{\strut ##1}}}}
\@namedef{PYG@tok@sh}{\def\PYG@tc##1{\textcolor[rgb]{0.87,0.13,0.00}{##1}}\def\PYG@bc##1{{\setlength{\fboxsep}{0pt}\colorbox[rgb]{1.00,0.94,0.94}{\strut ##1}}}}
\@namedef{PYG@tok@s1}{\def\PYG@tc##1{\textcolor[rgb]{0.87,0.13,0.00}{##1}}\def\PYG@bc##1{{\setlength{\fboxsep}{0pt}\colorbox[rgb]{1.00,0.94,0.94}{\strut ##1}}}}
\@namedef{PYG@tok@mb}{\let\PYG@bf=\textbf\def\PYG@tc##1{\textcolor[rgb]{0.00,0.00,0.87}{##1}}}
\@namedef{PYG@tok@mf}{\let\PYG@bf=\textbf\def\PYG@tc##1{\textcolor[rgb]{0.00,0.00,0.87}{##1}}}
\@namedef{PYG@tok@mh}{\let\PYG@bf=\textbf\def\PYG@tc##1{\textcolor[rgb]{0.00,0.00,0.87}{##1}}}
\@namedef{PYG@tok@mi}{\let\PYG@bf=\textbf\def\PYG@tc##1{\textcolor[rgb]{0.00,0.00,0.87}{##1}}}
\@namedef{PYG@tok@il}{\let\PYG@bf=\textbf\def\PYG@tc##1{\textcolor[rgb]{0.00,0.00,0.87}{##1}}}
\@namedef{PYG@tok@mo}{\let\PYG@bf=\textbf\def\PYG@tc##1{\textcolor[rgb]{0.00,0.00,0.87}{##1}}}
\@namedef{PYG@tok@ch}{\def\PYG@tc##1{\textcolor[rgb]{0.53,0.53,0.53}{##1}}}
\@namedef{PYG@tok@cm}{\def\PYG@tc##1{\textcolor[rgb]{0.53,0.53,0.53}{##1}}}
\@namedef{PYG@tok@cpf}{\def\PYG@tc##1{\textcolor[rgb]{0.53,0.53,0.53}{##1}}}
\@namedef{PYG@tok@c1}{\def\PYG@tc##1{\textcolor[rgb]{0.53,0.53,0.53}{##1}}}

\def\PYGZlt{\char`\<}
\def\PYGZgt{\char`\>}
\def\PYGZsh{\char`\#}

\def\PYGZhy{\char`\-}

\makeatother

\copyrightyear{2025}
\acmYear{2025}
\setcopyright{cc}
\setcctype{by}
\acmConference[SC Workshops '25]{Workshops of the International Conference for High Performance Computing, Networking, Storage and Analysis}{November 16--21, 2025}{St Louis, MO, USA}
\acmBooktitle{Workshops of the International Conference for High Performance Computing, Networking, Storage and Analysis (SC Workshops '25), November 16--21, 2025, St Louis, MO, USA}
\acmDOI{10.1145/3731599.3767358}
\acmISBN{979-8-4007-1871-7/2025/11}

\begin{document}
\title{An elastic job scheduler for HPC applications on the cloud}

\author{Aditya Bhosale}
\affiliation{%
  \institution{University of Illinois Urbana-Champaign}
  \city{Urbana, IL}
  \country{USA}}
\email{adityapb@illinois.edu}

\author{Kavitha Chandrasekar}
\affiliation{%
  \institution{University of Illinois Urbana-Champaign}
  \city{Urbana, IL}
  \country{USA}}
\email{kchndrs2@illinois.edu}

\author{Laxmikant Kale}
\affiliation{%
  \institution{University of Illinois Urbana-Champaign}
  \city{Urbana, IL}
  \country{USA}}
\email{kale@illinois.edu}

\author{Sara Kokkila-Schumacher}
\affiliation{%
 \institution{IBM Research}
 \city{Yorktown Heights}
 \state{NY}
 \country{USA}}
\email{saraks@ibm.com}

\renewcommand{\shortauthors}{Bhosale et al.}

\begin{abstract}
The last few years have seen an increase in adoption of the cloud for running HPC applications. The pay-as-you-go cost model of these cloud resources has necessitated the development of specialized programming models and schedulers for HPC jobs for efficient utilization of cloud resources. A key aspect of efficient utilization is the ability to rescale applications on the fly to maximize the utilization of cloud resources. Most commonly used parallel programming models like MPI have traditionally not supported autoscaling either in a cloud environment or on supercomputers. While more recent work has been done to implement this functionality in MPI, it is still nascent and requires additional programmer effort. Charm++ is a parallel programming model that natively supports dynamic rescaling through its migratable objects paradigm.
In this paper, we present a Kubernetes operator to run Charm++ applications on a Kubernetes cluster. We then present a priority-based elastic job scheduler that can dynamically rescale jobs based on the state of a Kubernetes cluster to maximize cluster utilization while minimizing response time for high-priority jobs. We show that our elastic scheduler, with the ability to rescale HPC jobs with minimal overhead, demonstrates significant performance improvements over traditional static schedulers.
\end{abstract}




\maketitle



\section{Introduction}

Cloud platforms have traditionally been used for embarrassingly parallel workloads and loosely coupled web services, microservices, etc. However, due to the cost-effective nature of cloud computing, there has been a steady increase in adoption by the HPC community. In recent years, rapid advancements in AI have only accelerated the adoption of cloud for HPC workloads.

The pay-as-you-go cost model of cloud resources has underscored the importance of efficient utilization of these resources. An important aspect for maximizing resource utilization is the ability to rescale applications on the fly.
Several popular distributed ML frameworks have evolved to support efficient execution in cloud environments, featuring well-developed modules for elastic training and fault tolerance~\cite{pytorch, ray}.

HPC faces additional challenges in running effectively on the cloud. First, HPC applications are typically strongly coupled and highly latency-sensitive. They need high-performance interconnects such as InfiniBand\textsuperscript{TM/SM} to show good strong scaling performance, which traditionally have not been available on cloud platforms. Recent developments, such as the Elastic Fabric Adapter (EFA)~\cite{aws-efa} in AWS and the increased availability of dedicated HPC instances with InfiniBand interconnects~\cite{azure-hpc}, have significantly improved network performance.

Second, HPC applications are not inherently fault-tolerant and cannot continue execution if one of the nodes is killed during execution. Thus, changing the resources assigned to a job at runtime requires careful handling by the application. As a result, most HPC frameworks do not support elastic execution.

Popular parallel programming frameworks such as MPI have not traditionally supported elastic execution. 
MPI applications requiring rescaling have in the past relied on checkpointing and restarting the application with a different number of ranks. This requires massive programming effort and application-specific knowledge to carefully handle checkpointing and restarting while maintaining correctness and balancing the load among ranks. Moreover, checkpointing to disk is an expensive operation, especially in a cloud environment.

MPI introduced support for dynamic rescaling of processes using the \texttt{MPI\_Comm\_Spawn} functions in MPI 2.x~\cite{mpi2}. However, not all MPI implementations have this functionality, and it requires a significant programming effort to support rescaling and efficient load balancing after rescaling. These functions are also primarily meant for increasing the number of ranks in an application and cannot directly handle scaling down the number of ranks. MPI Sessions~\cite{mpisessions}, introduced more recently in MPI 4.x, provides a new, more efficient framework for dynamic resource management in MPI jobs, which also handles scaling down the number of ranks. However, it does not address the huge programming effort required for implementing rescaling support in the application.

Charm++~\cite{b:kale-krishnan-charmpp, sc14charm} is an object-oriented parallel programming framework that natively supports rescaling and dynamic load balancing without additional programming effort~\cite{malleable2014}, making it an ideal choice for elastic execution.

As the use of cloud resources for running HPC applications increases, it becomes critical to have an efficient scheduler framework for managing these HPC jobs. This presents new challenges for job schedulers, such as batch scheduling, elastic scheduling, etc. There have been several projects in recent years aimed at these challenges. Some of these works are discussed in more detail in a later section of this paper. 
These projects have been able to make significant advances toward achieving HPC-Cloud convergence by developing scalable batch schedulers~\cite{volcano, flux, kueue}, demonstrating elastic scheduling for distributed ML training jobs~\cite{voda}, and demonstrating elastic execution of MPI jobs on a Kubernetes\textsuperscript{\textregistered} cluster~\cite{Kub2023}. However, the key challenge of developing an automated HPC job scheduler that can dynamically scale jobs up or down based on resource availability and job traffic conditions in a cloud environment remains unaddressed.

In this paper, we present an operator for running Charm++ applications on a Kubernetes cluster. We also present a priority-based elastic job scheduling policy that is integrated into the operator, which maximizes cluster utilization while minimizing response times and completion times for high-priority jobs by rescaling running jobs on the fly. We then study the scaling performance of Charm++ applications on a Kubernetes cluster. We also evaluate the effectiveness of our elastic scheduler by comparing it to other static scheduling strategies in varying job traffic conditions.

\section{Background}
\subsection{Charm++}

Charm++~\cite{b:kale-krishnan-charmpp, sc14charm} is an asynchronous message-driven parallel programming model. Users express computation in terms of objects (\textit{chares}) that communicate via remote method invocations. The mapping of these objects to physical processors is managed by the Charm++ runtime system. Each Processing Element (PE) runs a scheduler and has a message queue. The scheduler picks messages from the message queue and delivers them to the destination object.

When a remote method is called on an object, the destination PE of the object is looked up in a distributed location manager. The arguments to the remote method invocation are then serialized, packed into a message, and sent to the destination PE, where the message is enqueued in the message queue. The scheduler retrieves the message from the queue, deserializes it, and calls the method on the destination object with the deserialized arguments.

These chares are migratable and can be dynamically moved between processors by the runtime system. This allows the runtime system to support dynamic load balancing without additional programming effort by moving chares from overloaded processors to underloaded ones. To enable effective load balancing, Charm++ programs are often overdecomposed, i.e., they have more chares than processors.
Overdecomposition also results in additional performance improvements through computation-communication overlap, making Charm++ applications latency tolerant. This makes Charm++ a good fit for running on cloud platforms with high-latency interconnects~\cite{Bhosale2025Cloud}.

Charm++ is well-suited for dynamic applications with load imbalance and applications with frequent messaging between processes. Charm++ has been adopted in several large-scale applications such as molecular dynamics~\cite{namd2005, jain:isc2016}, computational cosmology~\cite{2007_ChaNGaScaling}, state space search~\cite{langerHiPC13stip}, discrete event simulation~\cite{mikidaPADS16}, and others.

\subsection{Shrink/Expand in Charm++}

\begin{figure}
    \centering
    \begin{subfigure}[b]{\linewidth}
        \centering
        \includegraphics[width=\linewidth]{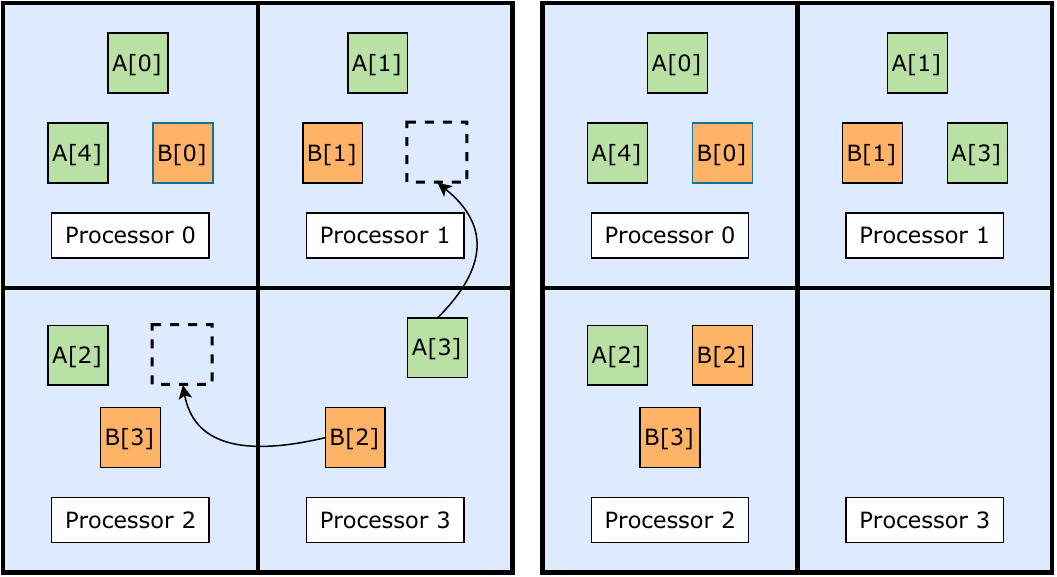}
        \caption{When shrinking from 4 to 3 PEs, the runtime moves objects away from processor 3}
        \label{fig:image1}
    \end{subfigure}
    
    \begin{subfigure}[b]{\linewidth}
        \centering
        \includegraphics[width=\linewidth]{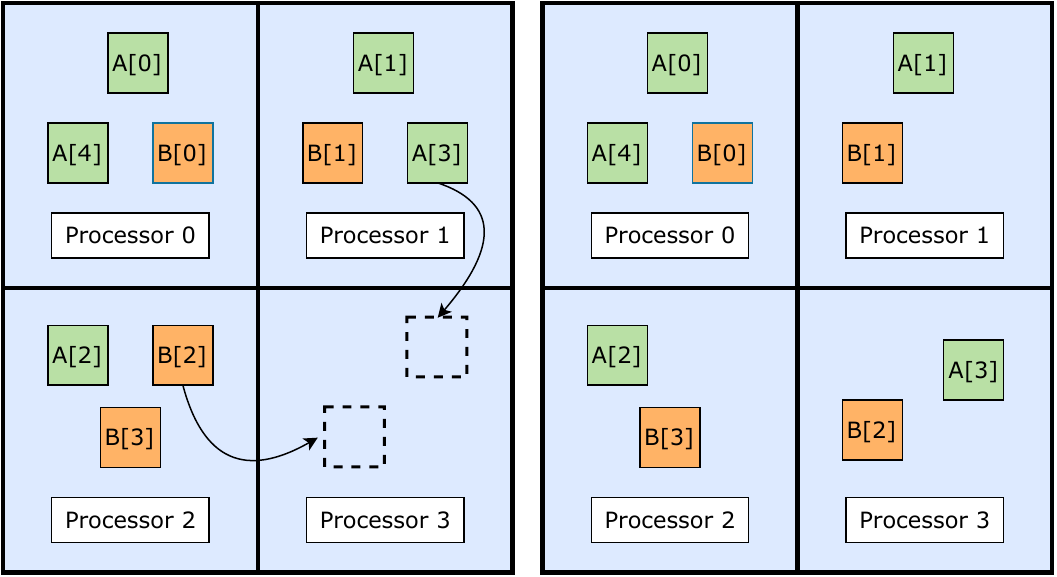}
        \caption{When expanding from 3 to 4 PEs, the load balancer moves objects from other PEs into the new processor to balance the load}
        \label{fig:image2}
    \end{subfigure}
    \caption{Shrink/Expand in Charm++}
    \label{fig:shrink-expand}
\end{figure}

Apart from dynamic load balancing, the concept of migratability discussed in the previous section also enables dynamic rescaling of resources at runtime~\cite{malleable2014}. To shrink the number of processes in the application, chares can be moved away from certain processors and redistributed among the remaining processors. Similarly, to expand the number of processes, objects from the existing processors can be redistributed to balance the load among the new processors. Figure~\ref{fig:shrink-expand} illustrates the shrink/expand functionality in Charm++. Rescaling is initiated by sending a signal to the Charm++ application from an external program using the Converse Client-Server (CCS) interface~\cite{charm-ccs}. The application then triggers rescaling during the next load-balancing step after receiving the signal.

On receiving a signal to \textit{shrink}, the load balancer in Charm++ disables the assignment of objects to the PEs to be removed. Thus, the load balancer moves objects out of the processes to be killed. To update the runtime data structures in Charm++ after rescaling, the application's state is checkpointed and the application is restarted with the new resources. The checkpointing is performed in Linux\textsuperscript{\textregistered} shared memory to avoid the high latency of reading from and writing to disk. Similarly, when a signal to \textit{expand} is received, the application is first restarted with the new processes. A load balancing step is performed after the restart to distribute the load evenly among the new processes.

Charm++ previously supported the rescaling functionality only in the \texttt{netlrts} build, which is a general-purpose, portable communication layer built on standard TCP/UDP sockets. As part of this work, we extended the rescaling functionality to the MPI build of Charm++, which resulted in a significant reduction in rescaling overheads.

\subsection{Kubernetes}

Kubernetes is an open-source software for cluster management and deployment of containerized applications. While initially designed for microservices and web applications, there has been a growing interest in using Kubernetes for HPC applications. Containerized execution on Kubernetes provides several advantages for HPC applications, such as reproducibility, consistent environments across different platforms, and ease of distribution. Kubernetes pod scheduling can be used for controlling CPU, memory, and GPU resources allocated to an HPC job. Other features, such as node affinity and anti-affinity, can be used for controlling pod placement to optimize communication performance.

While the default kube-scheduler doesn't support batch scheduling, more recent work has resulted in a rich ecosystem of efficient plugin schedulers developed specifically for HPC workloads~\cite{kube-batch, volcano, flux, Misale2021}.

The operator framework in Kubernetes can be used to extend the native Kubernetes functionality by implementing domain-specific application management features.
A Kubernetes operator consists of 2 main components. 
\begin{itemize}
    \item Custom Resource Definition (CRD) - which defines custom resource types for a specific application.
    \item Controller - A control loop that manages the custom resources and takes actions to maintain a desired state
\end{itemize}
Kubeflow's\textsuperscript{\textregistered} MPI Operator~\cite{MPIOperator} uses this operator pattern to manage MPI applications on Kubernetes. The MPI Operator launches a launcher pod and several worker replicas based on the MPI Job specification YAML file. Each worker replica can run several MPI ranks. Henceforth in this paper, job is used to indicate either an MPI Job or a Charm++ Job, and the use of replicas refers to the number of worker replicas in an MPI/Charm++ Job. The controller creates a hostfile with the domain names for worker pods used by MPI to connect to those workers. The worker pods run an SSH\textsuperscript{\textregistered} server and the launcher pod runs the \texttt{mpirun} command.

\section{Implementation}

\subsection{Charm++ operator}

To run Charm++ applications on Kubernetes clusters, we extended the MPI operator. We added functionality to rescale jobs when the deployment YAML file is modified. We use the default kube-scheduler for pod placement. We added pod affinity to the operator to ensure locality-aware placement of pods in the cluster. Similar to the hostfile, the controller creates a nodelist file that Charm++ uses to connect to the worker replicas.

We used the non-SMP build of Charm++ for our implementation, ie. each process is a single PE. We use a single process per worker replica to allow a more fine-grained control over elasticity.

Since Charm++ uses Linux shared memory for checkpointing, we do not need a persistent volume mount in our jobs. However, a common default shared memory size for each pod is 64MB. We used an \texttt{emptyDir} memory-backed volume mounted to \texttt{/dev/shm} to work around this restriction.

To shrink a running job, we follow these steps,
\begin{itemize}
    \item Send shrink signal to Charm++ application
    \item After the Charm++ application returns an acknowledgment for the shrink operation, remove extra pods from the job
\end{itemize}
Similarly, to expand a job,
\begin{itemize}
    \item Add new pods to a job
    \item Update the nodelist file to include new pods
    \item Send expand signal to the Charm++ application
\end{itemize}

\subsection{Elastic scheduling}

Previous work on a proof-of-concept adaptive scheduler for elastic Charm++ jobs on an HPC system has been shown to have significant performance improvements over scheduling rigid jobs for a small number of jobs~\cite{malleable2014}. This scheduler, however, did not consider user-defined priorities for jobs and used a First Come First Serve policy for job allocation, which can potentially result in under-utilization of the cluster if a job with large resource requirements blocks smaller jobs that were submitted later from filling the gaps in the cluster.
Our approach incorporates the following improvements beyond the previous work~\cite{malleable2014}: (a) considering user-defined priorities for making scheduling decisions, (b) out-of-order allocations if they improve cluster utilization, and (c) a Kubernetes operator that can handle a much larger number of jobs.

Several scheduling policies can utilize the rescaling functionality of Charm++ to maximize the cluster performance using different performance metrics. In this paper, we present and evaluate one such policy, which attempts to maximize cluster utilization while minimizing wait times for high-priority jobs. At the end of this section, we will discuss other factors that we haven't considered in our implementation, which may be of interest when defining an elastic scheduling policy.

\subsubsection{Scheduling policy}

We modified the MPI operator CRD to include \texttt{minReplicas} and \texttt{maxReplicas} fields for the workers specification. The scheduler can launch and rescale a job to any number of PEs between its minimum and maximum replicas configuration. We also added a \texttt{priority} field to the job specification. Two jobs with the same user-defined priority are prioritized based on the job submission time. The job that was submitted earlier has a higher priority. We do not modify the memory limits of the pods when rescaling a job. The memory limit in the worker specification is assumed to be the limit when running on the minimum replicas configuration.
To control the frequency of rescaling events, the scheduler maintains a configurable $T_{rescale\_gap}$ minimum gap between any two scheduling events (creation, shrink, expand)~\cite{malleable2014}.

Figure~\ref{code:scheduler:new} shows the pseudocode for the elastic scheduling policy when a new job is submitted. The scheduler scales down jobs with a lower priority as long as they meet the $T_{rescale\_gap}$ criteria until enough CPUs are freed to run the new job at its maximum replicas configuration. If the CPUs that would be freed by scaling down existing jobs are not enough to start the new job at its minimum replicas configuration, the new job is enqueued in an internal priority queue. 

\begin{figure}
\centering
\begin{Verbatim}[fontsize=\footnotesize, commandchars=\\\{\}]
\PYG{k}{def} \PYG{n+nf}{newJob}\PYG{p}{(}\PYG{n}{job}\PYG{p}{):}
    \PYG{n}{replicas} \PYG{o}{=} \PYG{n+nb}{min}\PYG{p}{(}\PYG{n}{freeSlots} \PYG{o}{\PYGZhy{}} \PYG{l+m+mi}{1}\PYG{p}{,} \PYG{n}{job}\PYG{o}{.}\PYG{n}{maxReplicas}\PYG{p}{)}
    \PYG{k}{if} \PYG{n}{replicas} \PYG{o}{\PYGZgt{}=} \PYG{n}{job}\PYG{o}{.}\PYG{n}{minReplicas}\PYG{p}{:}
        \PYG{n}{createOrExpandJob}\PYG{p}{(}\PYG{n}{job}\PYG{p}{,} \PYG{n}{replicas}\PYG{p}{)}
    \PYG{k}{else}\PYG{p}{:}
        \PYG{n}{numToFree} \PYG{o}{=} \PYG{n}{job}\PYG{o}{.}\PYG{n}{minReplicas} \PYG{o}{\PYGZhy{}} \PYG{n}{freeSlots} \PYG{o}{+} \PYG{l+m+mi}{1}
        \PYG{n}{index} \PYG{o}{=} \PYG{n+nb}{len}\PYG{p}{(}\PYG{n}{runningJobs}\PYG{p}{)} \PYG{o}{\PYGZhy{}} \PYG{l+m+mi}{1}
        \PYG{k}{while} \PYG{n}{numToFree} \PYG{o}{\PYGZgt{}} \PYG{l+m+mi}{0} \PYG{o+ow}{and} \PYG{n}{index} \PYG{o}{\PYGZgt{}} \PYG{l+m+mi}{0}\PYG{p}{:}
            \PYG{c+c1}{\PYGZsh{} runningJobs is a list of running jobs sorted}
            \PYG{c+c1}{\PYGZsh{} in decreasing order of priority}
            \PYG{n}{j} \PYG{o}{=} \PYG{n}{runningJobs}\PYG{p}{[}\PYG{n}{index}\PYG{o}{\PYGZhy{}\PYGZhy{}}\PYG{p}{]}
            \PYG{k}{if} \PYG{n}{currentTime}\PYG{p}{()} \PYG{o}{\PYGZhy{}} \PYG{n}{j}\PYG{o}{.}\PYG{n}{lastAction} \PYG{o}{\PYGZlt{}}
                \PYG{n}{rescaleGap}\PYG{p}{:}
                \PYG{k}{continue}
            \PYG{k}{if} \PYG{n}{j}\PYG{o}{.}\PYG{n}{priority} \PYG{o}{\PYGZgt{}} \PYG{n}{job}\PYG{o}{.}\PYG{n}{priority}\PYG{p}{:}
                \PYG{k}{break}
            \PYG{k}{if} \PYG{n}{j}\PYG{o}{.}\PYG{n}{replicas} \PYG{o}{\PYGZgt{}} \PYG{n}{j}\PYG{o}{.}\PYG{n}{minReplicas}\PYG{p}{:}
                \PYG{n}{newReplicas} \PYG{o}{=} \PYG{n+nb}{max}\PYG{p}{(}\PYG{n}{j}\PYG{o}{.}\PYG{n}{minReplicas}\PYG{p}{,}
                    \PYG{n}{j}\PYG{o}{.}\PYG{n}{replicas} \PYG{o}{\PYGZhy{}} \PYG{n}{numToFree}\PYG{p}{)}
                \PYG{n}{numToFree} \PYG{o}{\PYGZhy{}=} \PYG{p}{(}\PYG{n}{j}\PYG{o}{.}\PYG{n}{replicas} \PYG{o}{\PYGZhy{}}
                    \PYG{n}{newReplicas}\PYG{p}{)}

        \PYG{k}{if} \PYG{n}{numToFree} \PYG{o}{\PYGZgt{}} \PYG{l+m+mi}{0}\PYG{p}{:}
            \PYG{n}{enqueueJob}\PYG{p}{(}\PYG{n}{job}\PYG{p}{)}
            \PYG{k}{return}

        \PYG{n}{minToFree} \PYG{o}{=} \PYG{n}{job}\PYG{o}{.}\PYG{n}{minReplicas} \PYG{o}{\PYGZhy{}} \PYG{n}{freeSlots} \PYG{o}{+} \PYG{l+m+mi}{1}
        \PYG{n}{maxToFree} \PYG{o}{=} \PYG{n}{job}\PYG{o}{.}\PYG{n}{maxReplicas} \PYG{o}{\PYGZhy{}} \PYG{n}{freeSlots} \PYG{o}{+} \PYG{l+m+mi}{1}
        \PYG{n}{index} \PYG{o}{=} \PYG{n+nb}{len}\PYG{p}{(}\PYG{n}{runningJobs}\PYG{p}{)} \PYG{o}{\PYGZhy{}} \PYG{l+m+mi}{1}
        \PYG{k}{while} \PYG{n}{maxToFree} \PYG{o}{\PYGZgt{}} \PYG{l+m+mi}{0} \PYG{o+ow}{and} \PYG{n}{index} \PYG{o}{\PYGZgt{}} \PYG{l+m+mi}{0}\PYG{p}{:}
            \PYG{n}{j} \PYG{o}{=} \PYG{n}{runningJobs}\PYG{p}{[}\PYG{n}{index}\PYG{o}{\PYGZhy{}\PYGZhy{}}\PYG{p}{]}
            \PYG{k}{if} \PYG{n}{currentTime}\PYG{p}{()} \PYG{o}{\PYGZhy{}} \PYG{n}{j}\PYG{o}{.}\PYG{n}{lastAction} \PYG{o}{\PYGZlt{}}
                \PYG{n}{rescaleGap}\PYG{p}{:}
                \PYG{k}{continue}
            \PYG{k}{if} \PYG{n}{j}\PYG{o}{.}\PYG{n}{priority} \PYG{o}{\PYGZgt{}} \PYG{n}{job}\PYG{o}{.}\PYG{n}{priority}\PYG{p}{:}
                \PYG{k}{break}
            \PYG{k}{if} \PYG{n}{j}\PYG{o}{.}\PYG{n}{replicas} \PYG{o}{\PYGZgt{}} \PYG{n}{j}\PYG{o}{.}\PYG{n}{minReplicas}\PYG{p}{:}
                \PYG{n}{newReplicas} \PYG{o}{=} \PYG{n+nb}{max}\PYG{p}{(}\PYG{n}{j}\PYG{o}{.}\PYG{n}{minReplicas}\PYG{p}{,}
                    \PYG{n}{j}\PYG{o}{.}\PYG{n}{replicas} \PYG{o}{\PYGZhy{}} \PYG{n}{maxToFree}\PYG{p}{)}
                \PYG{n}{oldReplicas} \PYG{o}{=} \PYG{n}{j}\PYG{o}{.}\PYG{n}{replicas}
                \PYG{k}{if} \PYG{p}{(}\PYG{n}{shrinkJob}\PYG{p}{(}\PYG{n}{j}\PYG{p}{,} \PYG{n}{newReplicas}\PYG{p}{)):}
                    \PYG{n}{numFreed} \PYG{o}{=} \PYG{p}{(}\PYG{n}{oldReplicas} \PYG{o}{\PYGZhy{}}
                        \PYG{n}{newReplicas}\PYG{p}{)}
                    \PYG{n}{minToFree} \PYG{o}{\PYGZhy{}=} \PYG{n}{numFreed}
                    \PYG{n}{maxToFree} \PYG{o}{\PYGZhy{}=} \PYG{n}{numFreed}

        \PYG{k}{if} \PYG{n}{minToFree} \PYG{o}{\PYGZgt{}} \PYG{l+m+mi}{0}\PYG{p}{:}
            \PYG{n}{enqueueJob}\PYG{p}{(}\PYG{n}{job}\PYG{p}{)}
            \PYG{k}{return}

\end{Verbatim}
\caption{Pseudocode for priority-based scheduling algorithm when a new job is submitted}
\label{code:scheduler:new}
\end{figure}

Figure~\ref{code:scheduler:complete} shows the pseudocode for the elastic scheduling policy when a job finishes execution. The freed CPUs are reassigned to either existing running jobs or used to start new jobs from the internal job queue based on the priorities of these jobs.

Note that this algorithm tries to maximize cluster utilization by scaling jobs while using a minimal number of rescaling calls to avoid the rescaling overheads. Thus, it may not always ensure that the highest priority jobs are running with their maximum replicas. For instance, a lower priority job may be running with maximum replicas on the cluster when a higher priority job arrives. 
If the free slots in the cluster are not enough to run the higher priority job at its maximum replicas configuration, but are enough to run with its minimum replicas configuration, our scheduling algorithm will run the higher priority job at its minimum replicas configuration to avoid a shrink call to the lower priority job. However, if enough slots are not available to start the higher priority job even at its minimum replicas configuration, the lower priority job will be scaled down as long as the $T_{rescale\_gap}$ criteria is met to run the higher priority job.

\begin{figure}
\centering
\begin{Verbatim}[fontsize=\footnotesize, commandchars=\\\{\}]
\PYG{k}{def} \PYG{n+nf}{completeJob}\PYG{p}{(}\PYG{n}{job}\PYG{p}{):}
    \PYG{n}{numWorkers} \PYG{o}{=} \PYG{n}{freeWorkers}\PYG{p}{(}\PYG{n}{job}\PYG{p}{)}
    \PYG{c+c1}{\PYGZsh{} freeWorkers deletes the pods associated with the job}
    \PYG{c+c1}{\PYGZsh{} and returns the number of slots that were freed}
    \PYG{n}{index} \PYG{o}{=} \PYG{l+m+mi}{0}
    \PYG{k}{while} \PYG{n}{numWorkers} \PYG{o}{\PYGZgt{}} \PYG{l+m+mi}{0} \PYG{o+ow}{and} \PYG{n}{index} \PYG{o}{\PYGZlt{}} \PYG{n+nb}{len}\PYG{p}{(}\PYG{n}{allJobs}\PYG{p}{):}
        \PYG{c+c1}{\PYGZsh{} allJobs is a list of all running and queued jobs}
        \PYG{c+c1}{\PYGZsh{} sorted in decreasing order of priority}
        \PYG{n}{j} \PYG{o}{=} \PYG{n}{allJobs}\PYG{p}{[}\PYG{n}{index}\PYG{o}{++}\PYG{p}{]}
        \PYG{k}{if} \PYG{n}{currentTime}\PYG{p}{()} \PYG{o}{\PYGZhy{}} \PYG{n}{j}\PYG{o}{.}\PYG{n}{lastAction} \PYG{o}{\PYGZlt{}}
            \PYG{n}{rescaleGap}\PYG{p}{:}
            \PYG{k}{continue}
        \PYG{k}{if} \PYG{n}{j}\PYG{o}{.}\PYG{n}{replicas} \PYG{o}{\PYGZlt{}} \PYG{n}{j}\PYG{o}{.}\PYG{n}{maxReplicas}\PYG{p}{:}
            \PYG{n}{addReplicas} \PYG{o}{=} \PYG{n+nb}{min}\PYG{p}{(}\PYG{n}{numWorkers}\PYG{p}{,}
                \PYG{n}{j}\PYG{o}{.}\PYG{n}{maxReplicas} \PYG{o}{\PYGZhy{}} \PYG{n}{j}\PYG{o}{.}\PYG{n}{replicas}\PYG{p}{)}
            \PYG{k}{if} \PYG{n}{j}\PYG{o}{.}\PYG{n}{replicas} \PYG{o}{+} \PYG{n}{addReplicas} \PYG{o}{\PYGZgt{}=}
                \PYG{n}{j}\PYG{o}{.}\PYG{n}{minReplicas}\PYG{p}{:}
                \PYG{k}{if} \PYG{p}{(}\PYG{n}{createOrExpandJob}\PYG{p}{(}\PYG{n}{j}\PYG{p}{,}
                    \PYG{n}{j}\PYG{o}{.}\PYG{n}{replicas} \PYG{o}{+} \PYG{n}{addReplicas}\PYG{p}{)):}
                    \PYG{n}{numWorkers} \PYG{o}{\PYGZhy{}=} \PYG{n}{addReplicas}
    \PYG{n}{freeSlots} \PYG{o}{+=} \PYG{n}{numWorkers}
\end{Verbatim}
\caption{Pseudocode for priority-based scheduling algorithm when a job completes}
\label{code:scheduler:complete}
\end{figure}

\subsubsection{Discussion}

The following are some factors that we do not consider in the scheduling policy presented in this section, but that could be of interest for an elastic scheduler.

\paragraph{Fault tolerance}

Node failures are not an uncommon occurrence in cloud environments. In the scheduling policy we presented, we do not consider fault tolerance, as our goal was to demonstrate rescaling without requiring a shared filesystem. However, Charm++ natively supports fault tolerance by enabling checkpointing of chare data to disk every few iterations, and restarting from a checkpoint by adding an extra command-line parameter to the application launch command. The operator can be modified to launch with the extra restart parameter when a job restarts after a failure, which would start the application from the checkpoint if checkpoint data is found.

\paragraph{Job preemption}

The scheduling policy we presented does not preempt lower-priority jobs to run higher-priority jobs. This decision was again made to avoid the requirement of a shared filesystem. However, in the presence of a shared filesystem, lower-priority jobs could be sent a signal to checkpoint to disk and then be preempted to make room for higher-priority jobs. The lower-priority job can then be restarted from its checkpoint at a later time by simply using the restart command-line parameter.

\paragraph{Aging priorities}

A potential issue with the scheduling policy we presented is the starvation of low-priority jobs. A dynamic priority system could be implemented to gradually increase the priority of waiting jobs, ensuring that low-priority jobs get resources during times of high traffic.

\section{Evaluation}

For evaluation, we ran experiments on a 4-node Amazon EKS\textsuperscript{TM/SM} cluster with c6g.4xlarge instance type having 16 vCPUs each with a total of 64 vCPUs. We created all nodes in a single subnet in a cluster placement group for better networking performance. We used the non-SMP \texttt{mpi-linux-arm8} build of Charm++ with shrink/expand enabled, built with OpenMPI v3.1.

\subsection{Charm++ performance on Kubernetes}

First, we study the scaling performance of Charm++ applications on our EKS cluster using 2 applications.
\begin{itemize}
    \item Jacobi2D - This application solves the steady-state heat equation on a 2D grid using Jacobi iteration. This is a communication-intensive application.
    \item LeanMD - This is a molecular dynamics application that simulates atoms considering only the Lennard-Jones potential. This is a computationally intensive application.
\end{itemize}

\begin{figure}
    \centering
    \begin{subfigure}[b]{.45\linewidth}
        \centering
        \includegraphics[width=\linewidth]{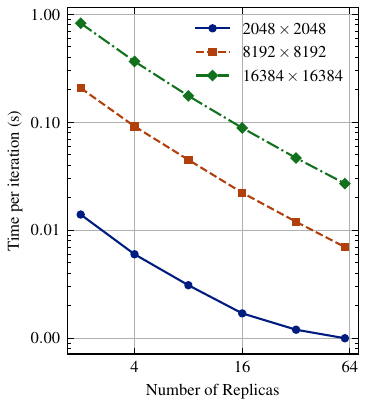}
        \caption{Strong scaling for Jacobi2D for different grid sizes}
        \label{fig:scaling:jacobi}
    \end{subfigure}
    \hfill
    \begin{subfigure}[b]{.45\linewidth}
        \centering
        \includegraphics[width=\linewidth]{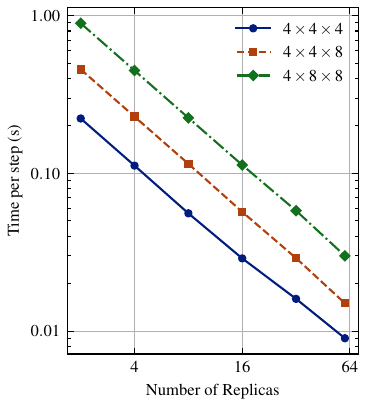}
        \caption{Strong scaling for LeanMD for different grid sizes}
        \label{fig:scaling:leanmd}
    \end{subfigure}
    \caption{Scaling performance of Charm++ on Kubernetes}
    \label{fig:scaling}
\end{figure}

Strong scaling results for Jacobi2D are seen in figure~\ref{fig:scaling:jacobi}. The different problem sizes are the dimensions of the 2D grid. We observe that for larger datasets, Jacobi2D scales well with an increasing number of replicas.

Figure~\ref{fig:scaling:leanmd} shows LeanMD's strong scaling performance. The problem sizes here are the number of cells in the application. Each cell has a fixed initial number of atoms. The simulation computes forces between atoms in the cells iteratively. Given that it's a compute-intensive application, we observe that LeanMD scales well with increasing replicas as expected.

\subsection{Rescaling overhead}

The rescaling overhead for Charm++ applications can be split into 4 parts~\cite{malleable2014}.
\begin{itemize}
    \item Checkpoint - Time taken to checkpoint state to Linux shared memory
    \item Restart - Time taken to restart the application
    \item Restore - Time taken to restore the checkpoint data from shared memory
    \item Load balance - Time taken to load balance. In the case of shrink, before the checkpoint/restart, and in the case of expand, after.
\end{itemize}
In this experiment, we use the 2D Jacobi solver from the previous section to study the contribution of each factor to the total rescaling overhead.

To study the effect of the number of replicas on the overhead, we use a constant grid size of $8k \times 8k$. We conduct 2 experiments. First, we shrink or scale down a job to half the number of replicas with a different number of original replicas and measure the time taken in each stage of rescaling. Second, we expand or scale up a job to double the number of replicas and take similar runtime measurements.

\begin{figure*}[h]
    \centering
    \begin{subfigure}[b]{.3\linewidth}
        \centering
        \includegraphics[width=\linewidth]{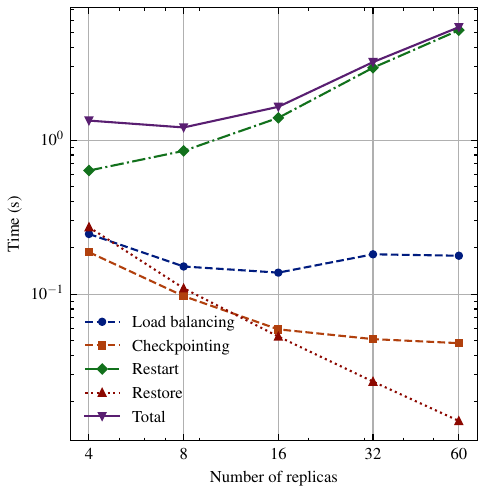}
        \caption{When shrinking to half the number of replicas. The x-axis shows the number of replicas before shrinking}
        \label{fig:overhead:shrink}
    \end{subfigure}
    \hfill
    \begin{subfigure}[b]{.3\linewidth}
        \centering
        \includegraphics[width=\linewidth]{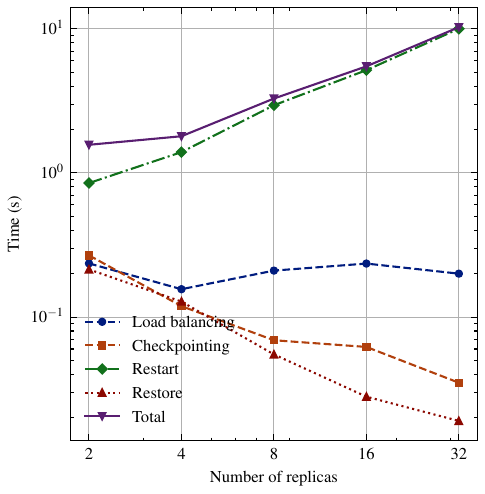}
        \caption{When expanding to double the number of replicas. The x-axis shows the number of replicas before expanding}
        \label{fig:overhead:expand}
    \end{subfigure}
    \hfill
    \begin{subfigure}[b]{.3\linewidth}
        \centering
        \includegraphics[width=\linewidth]{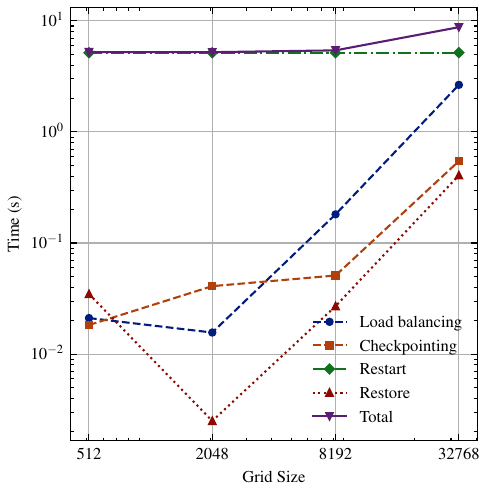}
        \caption{When shrinking from 32 to 16 replicas for different problem sizes. Grid size is the size of one dimension of the 2D grid}
        \label{fig:overhead:size}
    \end{subfigure}
    \caption{Contribution of different stages of rescaling to the total rescaling overhead}
    \label{fig:overhead}
\end{figure*}

Figure~\ref{fig:overhead:shrink} and~\ref{fig:overhead:expand} show the overhead of shrink and expand, respectively. We see that the application restart time increases with an increase in the number of replicas since the MPI startup time increases with the number of ranks. The checkpoint and restore time decreases with an increase in replicas because the size of the checkpoint data per replica reduces as we increase the number of replicas, since the total problem size is constant. The load balancing time stays flat with an increase in the number of replicas.

To study the effect of problem size on the overhead, we shrink jobs of different sizes from 32 to 16 replicas and measure the overhead of each stage. Figure~\ref{fig:overhead:size} shows the result of this experiment. We see that for small problem sizes, the overhead is dominated by the restart time. As the problem size increases, the load balancing, checkpoint, and restore time scale up while the restart time remains flat. We see that the overhead of in-memory checkpointing and restoring remains significantly low even for a problem with data size of 4GB.

\begin{figure}[h]
    \centering
    \begin{subfigure}[b]{.47\linewidth}
        \centering
        \includegraphics[width=\linewidth]{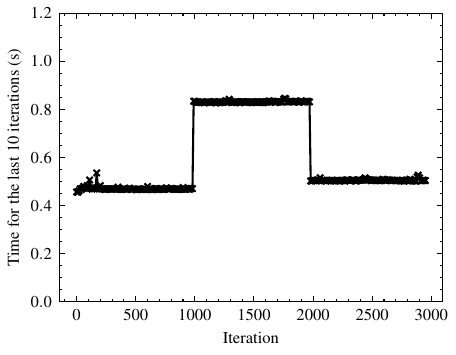}
        \caption{Time taken for the last 10 iterations}
        \label{fig:timeline:itertime}
    \end{subfigure}
    \hfill
    \begin{subfigure}[b]{.47\linewidth}
        \centering
        \includegraphics[width=\linewidth]{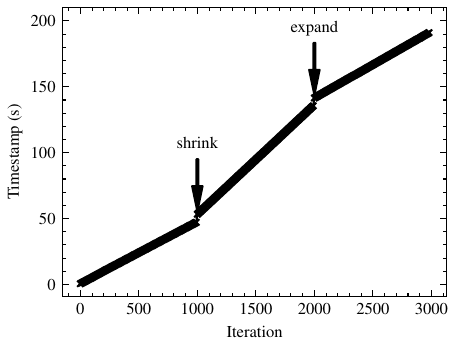}
        \caption{Timeline plot of every 10 iterations}
        \label{fig:timeline:overhead}
    \end{subfigure}
    \caption{Timeline for Jacobi2D with a \textit{shrink} from 32 replicas to 16 replicas and an \textit{expand} from 16 replicas to 32 replicas}
    \label{fig:timeline}
\end{figure}

Figure~\ref{fig:timeline:itertime} shows the time taken per 10 iterations for a $16k \times 16k$ grid. When the application is scaled down from 32 replicas to 16 replicas, we see that the time for each iteration increases. Similarly, when the application is scaled back up to 32 replicas, the time per iteration drops back down to its original value.

Figure~\ref{fig:timeline:overhead} shows the timeline plot for the time at which each iteration finishes execution. The slope of the line indicates the speed at which iterations are executed. We see that the slope of the line increases after scaling down, indicating a slower speed of executing iterations, and decreases after scaling up. The gaps in the timeline plot during shrink and expand show the overhead of rescaling.

\subsection{Elastic scheduling performance}

We compare our elastic scheduling policy with 3 other scheduling strategies. All these strategies use the same priority logic.
\begin{itemize}
    \item Rigid with \texttt{min\_replicas} - All jobs are launched with\\ \texttt{min\_replicas} number of replicas without rescaling
    \item Rigid with \texttt{max\_replicas} - All jobs are launched with\\ \texttt{max\_replicas} number of replicas without rescaling
    \item Moldable - The scheduler assigns the number of replicas based on the cluster state to maximize utilization, but doesn't rescale any job during execution~\cite{moldable1996}
\end{itemize}

We use the following metrics for evaluating the performance of these scheduling policies,
\begin{itemize}
    \item Total time - The end-to-end runtime from the start of the first job to the end of the last job
    \item Cluster utilization - Average percentage utilization of the cluster over the duration of the experiment
    \item Weighted mean response time - Mean response time weighted by job priority. We define response time as the time between a job submission and start
    \item Weighted mean completion time - Mean completion time weighted by job priority. Completion time is defined as time between a job submission and completion
\end{itemize}

The performance of the 4 schedulers depends on several factors, such as the size of the jobs, the order in which jobs are submitted, the rate of job submission, the choice of $T_{rescale\_gap}$, etc. An experimental study to measure the effect of each of these factors on the scheduler performance using actual runs would be infeasible. In this section we present the results from a simulator we wrote for modeling the scheduler performance for the Jacobi 2D example. We also present the results from an experimental run on a Kubernetes cluster.

\subsubsection{Simulation results} \label{sec:sim-results}

We use the Jacobi2D solver example from the previous section for our simulations.
We use 4 different problem sizes - small, medium, large, and xlarge based on their grid sizes and number of timesteps. 
\begin{itemize}
    \item Small - $512 \times 512$ grid. 40,000 timesteps. \texttt{min\_replicas} = 2, \texttt{max\_replicas} = 8
    \item Medium - $2048 \times 2048$ grid. 40,000 timesteps. \texttt{min\_replicas} = 4, \texttt{max\_replicas} = 16
    \item Large - $8192 \times 8192$ grid. 40,000 timesteps. \texttt{min\_replicas} = 8, \texttt{max\_replicas} = 32
    \item XLarge - $16,384 \times 16,384$ grid. 10,000 timesteps. \texttt{min\_replicas} = 16, \texttt{max\_replicas} = 64
\end{itemize}

We pick 16 jobs randomly out of these 4 sizes with random priorities between 1 and 5. We repeat this experiment 100 times and report the average metrics across all runs. We do not consider the overhead added by the operator or by Kubernetes to start up the pods.
We use strong scaling performance measurements for the 4 problem sizes to model the runtime of a job for a given number of replicas using a piecewise linear function. We also use the rescaling overhead measurements to model the overhead for rescaling using a piecewise linear function. 

\begin{figure*}
    \centering
    \begin{subfigure}[b]{.245\textwidth}
        \centering
        \includegraphics[width=\textwidth]{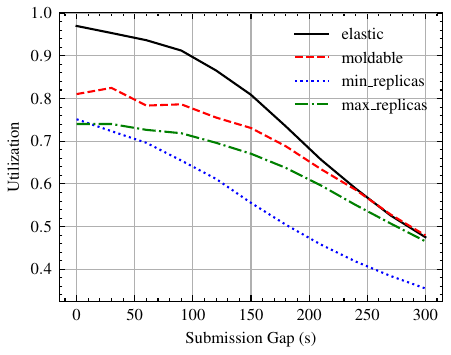}
        \caption{Cluster utilization vs submission gap}
        \label{fig:simulation_submission:utilization}
    \end{subfigure}
    \hfill
    \begin{subfigure}[b]{.245\textwidth}
        \centering
        \includegraphics[width=\textwidth]{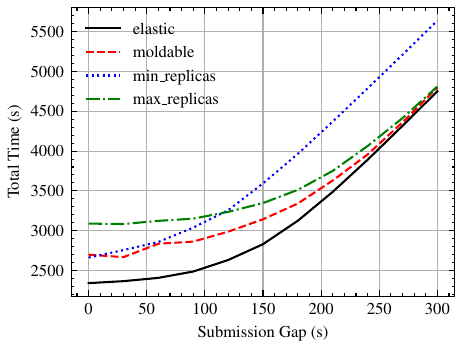}
        \caption{Total time vs submission gap\break}
        \label{fig:simulation_submission:total_time}
    \end{subfigure}
    \hfill
    \begin{subfigure}[b]{.245\textwidth}
        \centering
        \includegraphics[width=\textwidth]{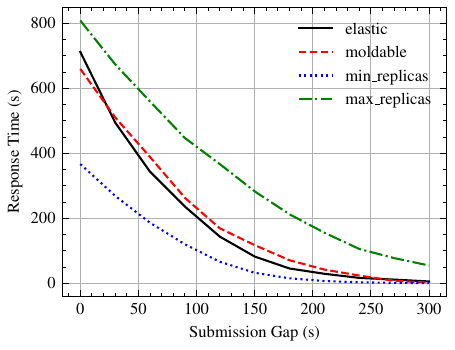}
        \caption{Weighted mean response time vs submission gap}
        \label{fig:simulation_submission:response_time}
    \end{subfigure}
    \hfill
    \begin{subfigure}[b]{.245\textwidth}
        \centering
        \includegraphics[width=\textwidth]{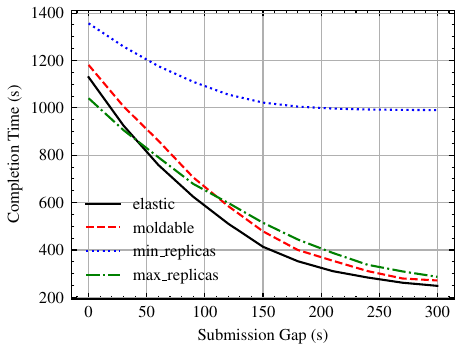}
        \caption{Weighted mean completion time vs submission gap}
        \label{fig:simulation_submission:completion_time}
    \end{subfigure}
    \caption{Results for scheduling performance simulation for different job submission rates. The submission gap is the time between two consecutive job submissions}
    \label{fig:simulation_submission}
\end{figure*}

\paragraph{Effect of Job Submission Rate}
We study the effect of job submission rate on the performance metrics defined earlier by varying the gap between two consecutive submissions from 0 to 300s. We used $T_{rescale\_gap} = 180s$ for these simulations. Figure~\ref{fig:simulation_submission} shows the results of the simulations.

The cluster utilization is highest for the elastic scheduler and lowest for the \texttt{min\_replicas} scheduler, as expected. The total time is lowest for the elastic scheduler. The \texttt{min\_replicas} scheduler has a lower total time than the \texttt{max\_replicas} scheduler when the submission gap is small because the jobs can run with a higher parallel efficiency when using a smaller number of replicas, and a small gap in job submissions results in a higher cluster utilization. As the submission gap is increased, however, the cluster utilization using the \texttt{min\_replicas} scheduler drops, and the higher parallel efficiency can no longer offset the lower utilization, which results in a larger total time. The total time for the other 3 schedulers converges as the submission gap increases, since with a large enough gap between jobs, each job can run to completion with the maximum number of replicas.

The weighted mean response time is lowest for the \texttt{min\_replicas} scheduler because the low cluster utilization leaves resources available for any higher priority jobs submitted later. The elastic scheduler has a better response time than the moldable and \texttt{max\_replicas} schedulers because of its ability to rescale low-priority jobs to run higher-priority jobs.

The weighted mean completion time is the highest for the\\ \texttt{min\_replicas} scheduler, even though the response time is low, because it runs jobs with a lesser degree of parallelism, thus increasing their runtime. The \texttt{max\_replicas} scheduler has the lowest weighted mean completion time for very small submission gaps because it can run the jobs in the correct priority order using the maximum degree of parallelism. The elastic and moldable schedulers have a slightly worse completion time for small submission gaps because they will run some jobs with a smaller number of replicas to fill the gaps in the cluster and maximize cluster utilization. As a result, the elastic and moldable schedulers have a higher cluster utilization and a lower total time and mean response time, but the increased runtime of these jobs increases the completion time.
As the submission gap increases, the \texttt{max\_replicas} and moldable schedulers can saturate the cluster with lower priority jobs before a higher priority job is submitted, resulting in worse completion times. The elastic scheduler has a lower mean completion time since it can rescale these running lower-priority jobs.

Our simulations show that the elastic scheduler can significantly reduce the end-to-end time of jobs and maximize cluster utilization while at the same time reducing the response time and completion time of jobs.

\begin{figure*}
    \centering
    \begin{subfigure}[b]{0.245\textwidth}
        \centering
        \includegraphics[width=\textwidth]{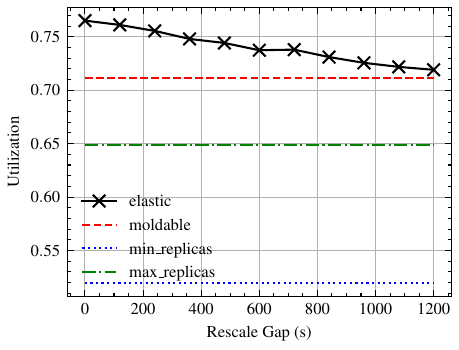}
        \caption{Cluster utilization vs $T_{rescale\_gap}$}
        \label{fig:simulation_rescale_gap:utilization}
    \end{subfigure}
    \hfill
    \begin{subfigure}[b]{0.245\textwidth}
        \centering
        \includegraphics[width=\textwidth]{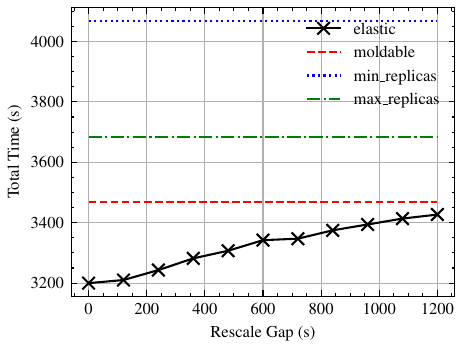}
        \caption{Total time vs $T_{rescale\_gap}$\break}
        \label{fig:simulation_rescale_gap:total_time}
    \end{subfigure}
    \hfill
    \begin{subfigure}[b]{0.245\textwidth}
        \centering
        \includegraphics[width=\textwidth]{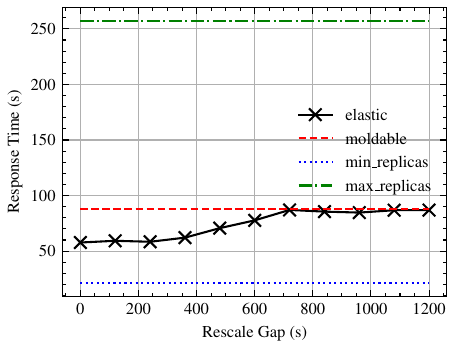}
        \caption{Weighted mean response time vs $T_{rescale\_gap}$}
        \label{fig:simulation_rescale_gap:response_time}
    \end{subfigure}
    \hfill
    \begin{subfigure}[b]{0.245\textwidth}
        \centering
        \includegraphics[width=\textwidth]{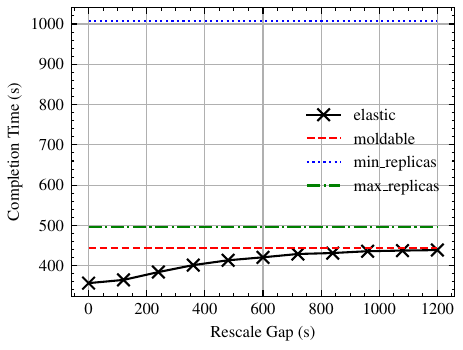}
        \caption{Weighted mean completion time vs $T_{rescale\_gap}$}
        \label{fig:simulation_rescale_gap:completion_time}
    \end{subfigure}
    \caption{Results for scheduling performance simulation for different $T_{rescale\_gap}$}
    \label{fig:simulation_rescale_gap}
\end{figure*}

\paragraph{Effect of $T_{rescale\_gap}$}
We repeat the previous experiment here with a fixed submission gap of $180s$ and vary $T_{rescale\_gap}$ from $0$ to $1200s$. Figure~\ref{fig:simulation_rescale_gap} shows the results of the simulations.

The cluster utilization is highest with a small $T_{rescale\_gap}$ because the scheduler can rescale running jobs more frequently to maximize utilization.
More frequent rescaling operations add a rescaling overhead to the total time.
However, figure~\ref{fig:simulation_rescale_gap:total_time} shows that the total time increases monotonically as the rescale gap is increased.
This is because the rescaling overhead is small enough that the effect of improved cluster utilization outweighs the penalty of a higher rescaling overhead.

The weighted mean response time and weighted completion time behave similarly to the total time. The mean response time for the \texttt{min\_replicas} scheduler is the lowest because of low cluster utilization. The low cluster utilization ensures that higher-priority jobs start early, but they take longer to complete. This is reflected in the higher mean completion time for the \texttt{min\_replicas} scheduler.
All the metrics for the elastic scheduler approach the moldable scheduler as $T_{rescale\_gap}$ is increased, since the moldable scheduler is essentially the elastic scheduler that never rescales any job.

These simulations show that the elastic scheduler performs well for a wide range of choices of $T_{rescale\_gap}$. They also show that the overhead of rescaling is low enough that a very small $T_{rescale\_gap}$ affects the performance of the elastic scheduler minimally.

\subsubsection{Experimental performance} \label{sec:exp-perf}

To experimentally measure and compare the performance of the schedulers on a Kubernetes cluster, we pick a configuration out of the randomly generated jobs by the simulator. We use $T_{rescale\_gap} = 180s$ and a job submission gap of 90s.
Since there is no load imbalance in this example, we only load balance when a job has to be rescaled.
The rigid job schedulers are emulated by setting the same value for \texttt{min\_replicas} and \texttt{max\_replicas} for all jobs. The moldable scheduler is emulated by setting a large $T_{rescale\_gap}$ value to prevent the jobs from rescaling after they are launched~\cite{malleable2014}.
Table~\ref{tab:scheduling-sim} shows the comparison of performance metrics obtained from the simulation shown in the previous section, along with the actual performance metrics obtained from the experimental run.

\begin{table*}
  \centering
  \renewcommand{\arraystretch}{1.5}
  \begin{tabular}{|p{2cm}|c|c|c|c|c|c|c|c|}
    \hline
    \multirow{2}{2cm}{\textbf{Scheduler}} & \multicolumn{2}{c|}{\thead{Total time (s)}} & \multicolumn{2}{c|}{\thead{Cluster\\utilization}} & \multicolumn{2}{c|}{\thead{Weighted mean\\response time (s)}} & \multicolumn{2}{c|}{\thead{Weighted mean\\completion time (s)}}\\
    \cline{2-9}
    & \textbf{Actual} & \textbf{Simulation} & \textbf{Actual} & \textbf{Simulation} & \textbf{Actual} & \textbf{Simulation} & \textbf{Actual} & \textbf{Simulation}\\
    \hline
    \texttt{min\_replicas} & 2511 & 2402 & 58.12\% & 60.88\% & 162.87 & 207.21 & 888.59 & 915.08 \\
    \texttt{max\_replicas} & 2149 & 1914 & 81.32\% & 85.86\% & 147.51 & 195.79 & 275.36 & 326.68\\
    Moldable & 2111 & 2078 & 71.54\% & 78.39\% & 92.93 & 122.40 & 273.89 & 326.15 \\
    Elastic & 1796 & 1813 & 87.80\% & 92.26\% & 69.93 & 32.96 & 265.76 & 241.29  \\ \hline
  \end{tabular}
  \caption{Actual and simulation performance results for the 4 scheduling policies}
  \label{tab:scheduling-sim}
\end{table*}

The \texttt{min\_replicas} scheduler has the lowest cluster utilization since all jobs run with the minimum number of replicas specified by the user. When running with \texttt{max\_replicas}, the utilization is higher, but because of the rigidity of jobs, some gaps in utilization are not filled. 
The total time for the \texttt{max\_replicas} scheduler is lower than the total time for the \texttt{min\_replicas} scheduler because of the low cluster utilization of the latter.
Both the \texttt{min\_replicas} and \texttt{max\_replicas} schedulers have a high response time since higher-priority jobs that were submitted after the cluster was fully utilized had to wait for the lower-priority jobs to finish executing.

\begin{figure}
    \centering
    \begin{subfigure}[b]{\linewidth}
        \centering
        \includegraphics[width=\linewidth]{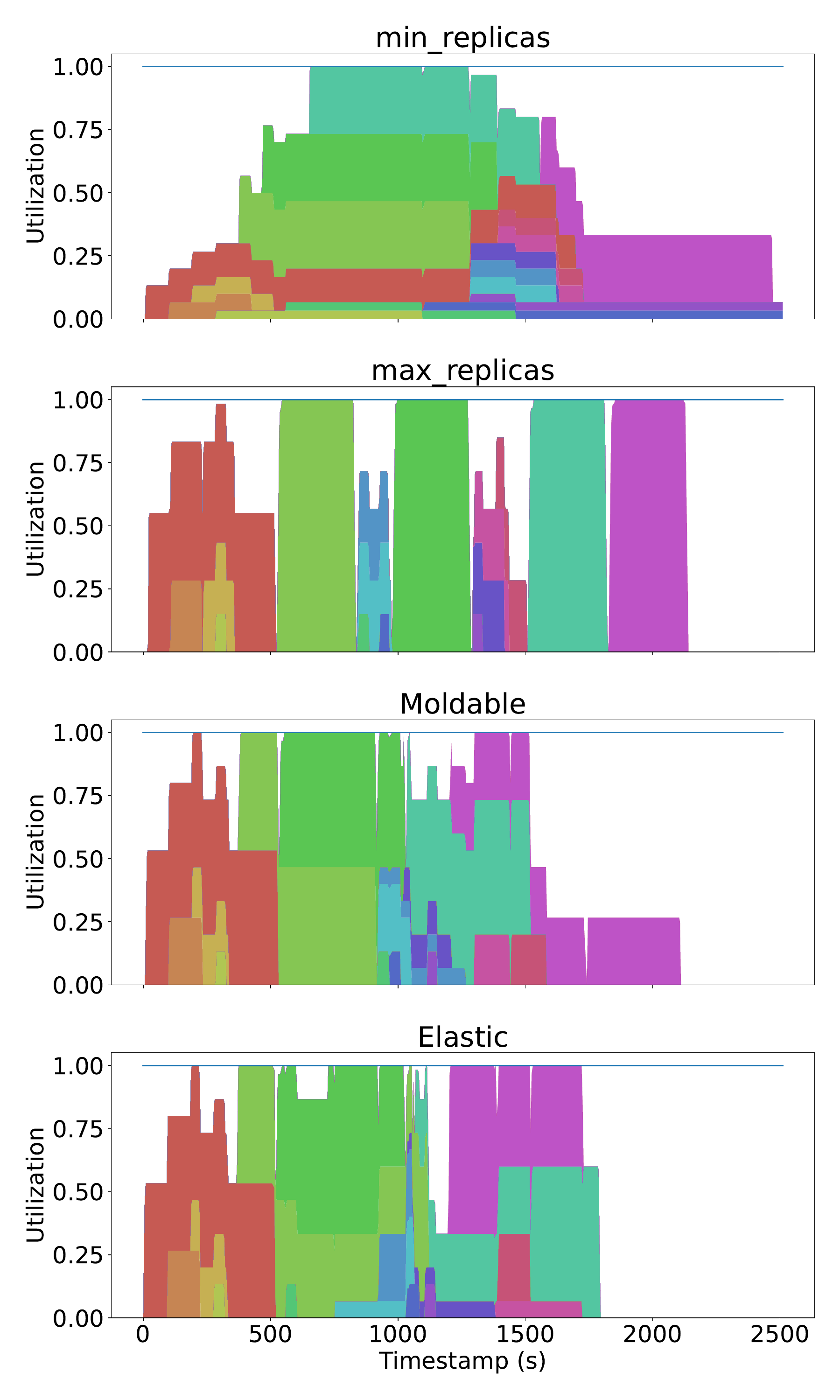}
        \caption{Cluster utilization profiles for the 4 schedulers we study in this experiment. Each color represents a different job}
        \label{fig:scheduling:utilization}
    \end{subfigure}
    
    \vspace{0.4cm}
    \begin{subfigure}[b]{\linewidth}
        \centering
        \includegraphics[width=\linewidth]{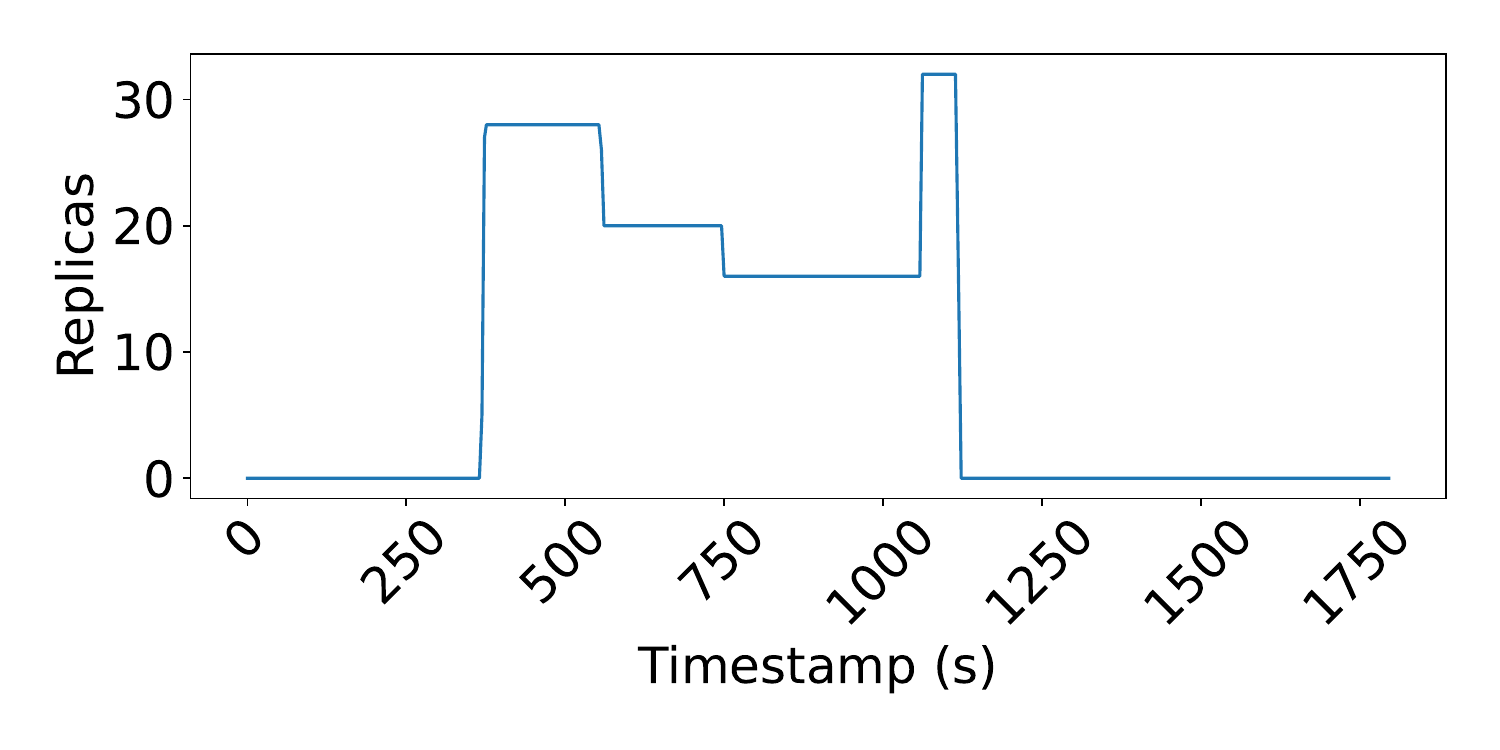}
        \caption{Evolution of number of replicas with time for an xlarge job using the elastic scheduler}
        \label{fig:scheduling:example}
    \end{subfigure}
    \caption{Results for scheduling performance comparison experiments on Amazon EKS}
    \label{fig:scheduling}
\end{figure}

In the case of moldable scheduling, the cluster utilization is high because the scheduler always picks the number of replicas to maximize cluster utilization. However, because the jobs cannot be rescaled after they are launched, a job that was submitted during high traffic can continue to run at a lower configuration even after the cluster has free slots, as can be seen in our experiment in figure~\ref{fig:scheduling:utilization}. The response time for the moldable scheduler is lower than the \texttt{max\_replicas} scheduler even when they have similar utilization because the moldable scheduler is able to start higher-priority jobs at a lower number of replicas configuration earlier, unlike the \texttt{max\_replicas} scheduler that has to queue them until the required slots are free.

The elastic scheduler is able to maximize the cluster utilization as well as minimize the mean response time and the mean completion time. The increase in cluster utilization compared to other scheduling strategies can be seen in figure~\ref{fig:scheduling:utilization}. When a higher-priority job is submitted to the cluster, unlike the moldable scheduler, the elastic scheduler can rescale the running lower-priority jobs to make room for the higher-priority job. Figure~\ref{fig:scheduling:example} shows an xlarge job that rescales multiple times with the elastic scheduler.

\section{Related work}

Over the past few years, there have been several projects to enable efficient execution of HPC workloads on the cloud. Previous work on running HPC applications on Kubernetes broadly follows two approaches. The first is to develop an efficient scheduler for controlling pod placement to minimize communication cost for MPI applications. The second is to develop a higher-level operator focusing on domain-specific management of applications.

Volcano\textsuperscript{TM} scheduler~\cite{volcano} is a Kubernetes-native batch scheduling system for HPC workloads. It supports features like network topology-aware scheduling, priority-based scheduling, cross-cluster job scheduling, etc. The MPI operator has support for using Volcano for gang scheduling of pods.

Kubeflux~\cite{Misale2021} is a plugin scheduler for Kubernetes for running HPC workloads. It demonstrated better scaling performance than the default kube-scheduler using the MPI operator.

More recent work on Kubeflux, now called Fluence, demonstrates the scalability of the scheduler by running production-grade MPI applications on up to 3000 ranks~\cite{flux}. The authors implemented several optimizations in the MPI operator to improve scalability and reliability. They demonstrated large-scale runs of production-grade MPI applications with good strong scaling performance and low variability.

Kueue~\cite{kueue} is a job queueing and resource management system for managing batch workloads. They support various types of jobs, including Kubeflow's MPIJob. Kueue also supports other features such as partial admission with reduced parallelism based on user quota, topology-aware scheduling, etc.

Voda scheduler~\cite{voda} extends the MPI operator for elastic distributed training. It supports locality-aware pod placement and addresses the issue of fragmented allocation after rescaling by supporting worker migration.

Kub~\cite{Kub2023} is a Kubernetes-based framework for elastic execution of MPI applications on the cloud. The authors demonstrate running MPI applications on the cloud with rescaling by checkpointing and restarting the application. However, they do not consider resource availability and use a static scaling protocol while making scaling decisions.

\section{Future Work}

The elastic scheduler presented in this paper has some limitations that we plan to address, and presents several avenues to explore in the future.
When making scheduling decisions, we do not consider the cost versus the potential benefit of rescaling a job. The potential benefit of rescaling depends on several factors,

\begin{itemize}
\item \textit{Number of replicas by which a job is scaling up}. A small increase in the number of replicas may not justify the overhead of rescaling.
\item \textit{Percentage of job already completed}. If only a small fraction of a job remains, scaling up may not provide enough benefit. Similarly, allowing the job to complete would be more efficient than scaling it down to start another job.
\item \textit{Parallel efficiency of the job}. If the job has a very low parallel efficiency at its current number of replicas, scaling up may not improve performance enough to offset the rescaling cost.
\end{itemize}

Incorporating some of these factors into the scheduler's decision-making requires input from the application during the decision-making process. This could be achieved by giving the application control to accept or decline a rescaling command sent by the scheduler. The application can then decline a rescaling command if only a small fraction of the application run remains. It can also decline a scaling-up command if the parallel efficiency of the job, as measured by runtime instrumentation, is lower than a specified threshold.

Another avenue to explore is running Charm++ applications as evolving jobs. Unlike elastic jobs, where the rescaling signal is sent from an external scheduler, evolving jobs can rescale at runtime based on internal, application-specific criteria without any external trigger~\cite{moldable1996}. This approach could be particularly useful in applications where the compute load changes dynamically over time, for example, due to dynamic refinement of the underlying grid in a numerical solver.

Charm++ also supports execution on GPUs~\cite{HAPI}. However, shrink/expand operations on GPUs aren't supported yet. In the future, we plan to support rescaling with GPUs and extend our operator to manage GPU resources. This will enable us to use the operator for running elastic distributed ML training workloads, in addition to other large-scale HPC applications that already use the GPU support in Charm++~\cite{2009ChaNGaGPU, namd2020}.

\section{Conclusion}

In this paper, we extended the KubeFlow MPI operator to run Charm++ applications on a Kubernetes cluster. We added functionality for dynamic rescaling of jobs to the operator. We showed scaling performance results for two Charm++ applications. We also measured the overhead of rescaling operations and attributed the cost to different stages of rescaling. We found that the overhead is dominated by the restart time for a problem size of up to 4GB, with a very low overhead of in-memory checkpointing and restoring.

We presented a priority-based elastic job scheduling policy that can dynamically rescale jobs based on the state of the cluster to maximize utilization and minimize response times for high-priority jobs.

We evaluated the performance of the elastic scheduling policy against two rigid scheduling policies and a moldable scheduling policy on 4 different performance metrics. We wrote a scheduling simulator to compare the performance of the 4 scheduling policies across a wide range of job submission patterns and showed that the elastic scheduler performs better than the other 3 scheduling policies. We ran an example job set on an AWS EKS Kubernetes cluster and showed that the elastic scheduler performed better on all 4 performance metrics.

We showed that traditional rigid job scheduling policies can perform well under certain conditions - the \texttt{min\_replicas} scheduler performs well under high-traffic conditions, while the \texttt{max\_replicas} scheduler performs well in low-traffic conditions. The dynamic scheduling policies, i.e., elastic and moldable, show good performance in all traffic conditions, with the elastic scheduling policy outperforming the moldable scheduling policy on all metrics.

\begin{acks}

This work is supported by the IBM-Illinois Discovery
Accelerator Institute (IIDAI).
\end{acks}

\bibliographystyle{ACM-Reference-Format}
\bibliography{citations, group} 

\clearpage

\twocolumn[%
{\begin{center}
\Huge
Appendix: Artifact Description/Artifact Evaluation
\end{center}}
]

\appendixAD

\section{Overview of Contributions and Artifacts}

\subsection{Paper's Main Contributions}

\begin{enumerate}[label=$C_{\arabic*}$]
    \item A Charm++ build supporting shrink/expand with the MPI communication layer.
    \item A modified MPI Operator implementation that can run Charm++ applications on a Kubernetes cluster with the 4 scheduling policies described in the paper.
    \item A simulator to model the performance of the 4 schedulers studied in this paper.
\end{enumerate}

\subsection{Computational Artifacts}

\begin{enumerate}[label=$A_{\arabic*}$]
    \item \url{https://github.com/adityapb/mpi-operator/tree/aws}
    \item \url{https://github.com/adityapb/mpi-operator/tree/aws/examples/v2beta1/charm/simulation}
\end{enumerate}

\begin{tabular}{|l|c|r|}
\hline
Artifact ID & Contribution supported & Related Paper Elements \\
\hline
$A_1$         & $C_2$         & Figure~\ref{fig:scheduling} \& Table~\ref{tab:scheduling-sim}            \\
$A_2$         & $C_3$         & Figures~\ref{fig:simulation_submission},~\ref{fig:simulation_rescale_gap} \& Table~\ref{tab:scheduling-sim}            \\
\hline
\end{tabular}

\section{Artifact Identification}

\newartifact

This artifact contains the source code for our modifications to the MPI Operator for running Charm++ applications with the elastic scheduling policy described in the paper. The following are the steps to run the experiment described in Section~\ref{sec:exp-perf},

\begin{enumerate}
    \item Set up an AWS EKS cluster with a single node group consisting of 4 c6g.4xlarge instances in a single subnet and a cluster placement group.
    \item To deploy the elastic scheduling policy, edit line 8244 in the file deploy/v2beta/mpi-operator.yaml to point to the image \texttt{adityapb/mpi-operator:mpi}
    \item Deploy the MPI Operator using the command -\\
    \texttt{kubectl create -f deploy/v2beta/mpi-operator.yaml}
    \item Generate the job YAML files by running the commands,\\
    \texttt{cd examples/v2beta1/charm}\\
    \texttt{python generate\_jobs.py generate elastic}
    \item Start the script for monitoring the cluster utilization by running,\\
    \texttt{python track\_utilization.py elastic}
    \item Submit the jobs to the cluster by running,\\
    \texttt{python generate\_jobs.py submit elastic}
    \item After all the jobs finish running, interrupt the utilization monitoring script. It will create a file called\\ \texttt{pod\_utilization\_elastic.log}
    \item To deploy the moldable scheduling policy, edit line 8244 in the file deploy/v2beta/mpi-operator.yaml to point to the image \texttt{adityapb/mpi-operator:moldable}
    \item Repeat steps 3 to 7 by replacing \texttt{elastic} with \texttt{moldable} in all commands to generate the file\\ \texttt{pod\_utilization\_moldable.log}
    \item Using the same MPI operator deployment, get the results for the other 2 schedulers by replacing \texttt{elastic} with \texttt{min} and \texttt{max}
    \item Run the following script to generate Figure~\ref{fig:scheduling} and print the actual metrics presented in Table~\ref{tab:scheduling-sim},\\
    \texttt{python plot\_utilization.py}
\end{enumerate}

The estimated execution time for the jobs of each scheduler to finish is around 50 minutes. For the 4 scheduling policies, the total execution time will be close to 4 hours.

\newartifact

This artifact contains the source code for the simulator to model the performance of the 4 scheduling policies studied in this paper. The following are the steps to reproduce the experiments presented in Section~\ref{sec:sim-results},

\begin{enumerate}
    \item Run the following command to generate the plots presented in Figures~\ref{fig:simulation_submission} and~\ref{fig:simulation_rescale_gap}, and print the simulation metrics presented in Table~\ref{tab:scheduling-sim},\\
    \texttt{cd examples/v2beta1/charm/simulation}\\
    \texttt{python run.py}
\end{enumerate}

The total execution time for the simulator should be less than 5 minutes.

\end{document}